\documentclass[%
 reprint,
superscriptaddress,
 amsmath,amssymb,
prb,
]{revtex4}

\usepackage{graphicx}% Include figure files
\usepackage{dcolumn}% Align table columns on decimal point
\usepackage{bm}% bold math
\usepackage{hyperref}% add hypertext capabilities
\usepackage[group-separator={,}]{siunitx}   % SI units
\begin{document}

\preprint{APS/123-QED}

\title{Distinct Ultrafast Electronic and Magnetic Response in M-edge Magnetic Circular Dichroism}% Force line breaks with \\

\author{Kelvin Yao}
\author{Felix Willems}%
\author{Clemens von Korff Schmising}
 \email{korff@mbi-berlin.de}
 \author{Ilie Radu}
 \thanks{present address: Department of Physics, Free University Berlin, Arnimallee 14, 14195 Berlin, Germany }
  \affiliation{%
 Max-Born-Institut für Nichtlineare Optik und Kurzzeitspektroskopie, Max-Born-Straße 2A, 12489 Berlin, Germany
}%
 \author{Christian Strüber}
 \thanks{present address: Department of Physics, Free University Berlin, Arnimallee 14, 14195 Berlin, Germany }
  \affiliation{%
 Max-Born-Institut für Nichtlineare Optik und Kurzzeitspektroskopie, Max-Born-Straße 2A, 12489 Berlin, Germany
}%
\author{Daniel Schick}
 \author{Dieter Engel}
   \affiliation{%
 Max-Born-Institut für Nichtlineare Optik und Kurzzeitspektroskopie, Max-Born-Straße 2A, 12489 Berlin, Germany
}%
 \author{Arata Tsukamoto}
 \affiliation{College of Science and Technology, Nihon University, 7-24-1 Funabashi, Chiba, Japan}

 \author{J. K. Dewhurst}
 \affiliation{Max-Planck-Institut für Mikrostrukturphysik, Weinberg 2, D-06120 Halle, Germany}
 \author{Sangeeta Sharma}
 \affiliation{%
 Max-Born-Institut für Nichtlineare Optik und Kurzzeitspektroskopie, Max-Born-Straße 2A, 12489 Berlin, Germany
}%
 \author{Stefan Eisebitt}
\affiliation{%
 Max-Born-Institut für Nichtlineare Optik und Kurzzeitspektroskopie, Max-Born-Straße 2A, 12489 Berlin, Germany
}%
\affiliation{Technische Universität Berlin, Institut für Optik und Atomare Physik, 10623 Berlin, Germany}

\date{\today}% It is always \today, today,

\begin{abstract}
Experimental investigations of ultrafast magnetization dynamics increasingly employ resonant magnetic spectroscopy in the ultraviolet spectral range. Besides allowing to disentangle the element-specific transient response of functional magnetic systems, these techniques also promise to access attosecond to few-femtosecond dynamics of spin excitations. Here, we report on a systematic study of transient magnetic circular dichroism (MCD) on the transition metals Fe, Co and Ni as well as on a FeNi and GdFe alloy and reveal a delayed onset between the electronic and magnetic response. Supported by \textit{ab-initio} calculations, we attribute our observation to a transient energy shift of the absorption and MCD spectra at the corresponding elemental resonances due to non-equilibrium changes of electron occupations.

\end{abstract}

\maketitle

%\tableofcontents

\section{\label{sec:Introduction}Introduction:\protect }

The validity and significance of experimental observables in ultrafast magnetization dynamics has been a matter of intense debate ever since it was first discovered that ultrashort optical pulse excitation leads to a magnetic response on the femtosecond time scale \cite{Beaurepaire1996}. Femtosecond Kerr or Faraday techniques in the visible spectral range may be influenced by non-equilibrium charge distributions and state-filling effects leading to complicated magneto-optical response functions that no longer reflect genuine spin excitations \cite{Koopmans2000,Zhang2009,Carva2011}. Nevertheless, it is generally accepted that by measuring the Voigt vector, i.e. the complete set of magnetic and non-magnetic quantities, in conjunction with a careful analysis and theoretical modeling, ultrafast magnetization dynamics can be determined with optical spectroscopy even on very early time scales \cite{Shokeen2017a}.
With the advancements in the field of femtomagnetism, the focus has shifted to more complex magnetic materials, where a light driven functionality may be related to interface effects \cite{Kampfrath2013} or to the direct interplay of different elements in nanostructures or multi-component systems \cite{Stanciu2007,Radu2011,Dewhurst2018b,Hofherr2020,Tengdin2020a,Willems2020}. Understanding these phenomena requires to retrieve an \textit{element-specific} response and has led to great efforts worldwide to develop time-resolved magnetic circular dichroism (MCD) techniques in the soft X-ray and extreme ultraviolet spectral (XUV) range. Such experiments have been pioneered at slicing sources of synchrotron facilities \cite{Stamm2007a}, free-electron lasers (FEL) \cite{Gutt2010} and laboratory-based high harmonic generation (HHG) radiation sources \cite{La-O-Vorakiat2009,Willems2015}. The potential to probe magnetization dynamics on a nanometer spatial scale \cite{Pfau2012,Vodungbo2012a,Graves2013,VonKorffSchmising2014} as well as to access attosecond to few femtosecond timescales \cite{Siegrist2019} are further important motivations to improve these short-wavelength based techniques.  
Again, a considerable effort is currently underway to characterize the experimental observables and to quantify possible electronic contributions in XUV spectroscopy \cite{La-O-Vorakiat2012} as well as to find theoretical descriptions for the resonant magnetic response in single \cite{Oppeneer2004,Turgut2016,Zusin2018a,Jana2020} and multi-element magnetic systems \cite{Dewhurst2020}. 

In this Rapid Communication, we report on helicity dependent transient absorption and MCD measurements on the transition metals Fe, Co and Ni as well as on GdFe and NiFe alloys and experimentally demonstrate a significantly delayed response of the MCD signal with respect to the onset of changes in the absorption, i.e., of the electronic system. This delay is different for the different elements, approximately \SI{55}{fs} for Ni, \SI{45}{fs} for Fe and not present for Co. We argue that the delayed onset of the transient magnetization signal is not caused by the evolution of the magnetic moment itself, but instead, discuss this observation in terms of a possible laser-driven ultrafast shift of the electronic and magnetic elemental absorption resonances towards lower photon energies. With a probing pulse of a finite spectral bandwidth, as often employed in HHG and FEL measurements, we can rationalize how a relative delay emerges by the interplay of two competing processes. On the one hand, a shift can result in an increase of the MCD signal while on the other hand, ultrafast demagnetization reduces the MCD signal. This assumption implies that the absolute time interval observed as a delay will depend sensitively on the spectral width and position of the probing XUV pulse as well as on the relative magnitude of the two transient contributions. 

\section{\label{sec:Experiment}Experiment:\protect}
\begin{figure}
\includegraphics{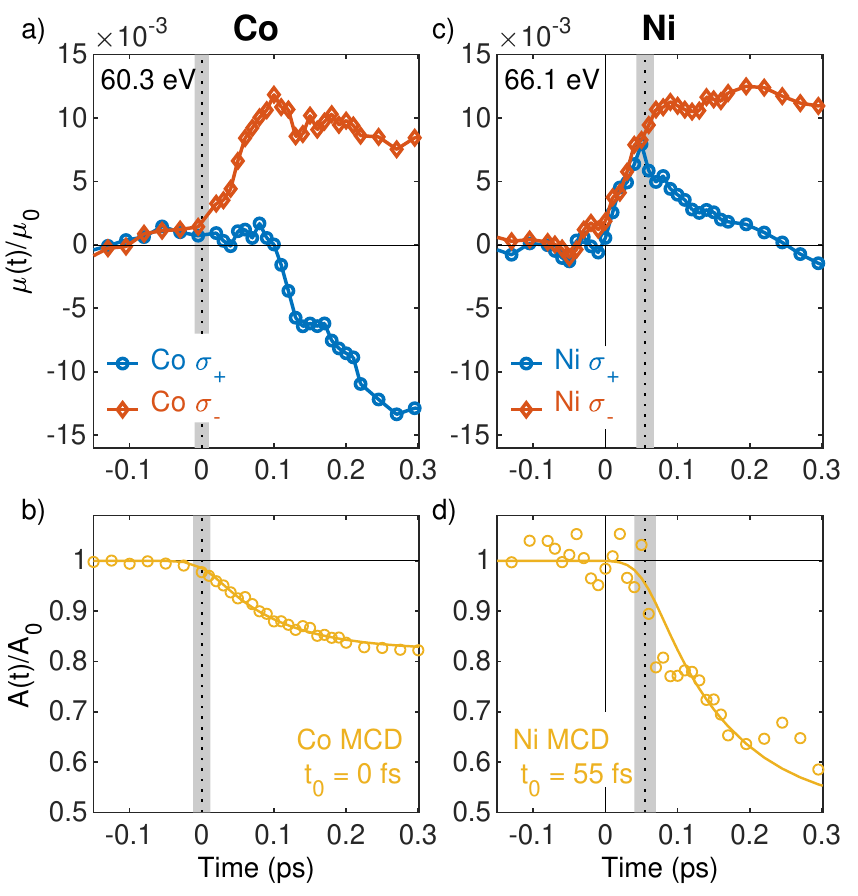}% Here is how to import EPS art
\caption{Normalized helicity dependent absorption $\mu(t)/\mu_0$ for a) Co at \SI{60.3}{eV} and c) Ni at \SI{66.1}{eV}. The corresponding normalized asymmetry $A(t)/A_0$ at these photon energies of Co and Ni are shown in b) and d), respectively. Note the apparent delay between the onset of absorption and magnetic changes in Ni of approximately $(55\pm12)\,\si{fs}$, marked as grey bars.}
\label{fig:CoNi} 
\end{figure}

\begin{figure}
\includegraphics{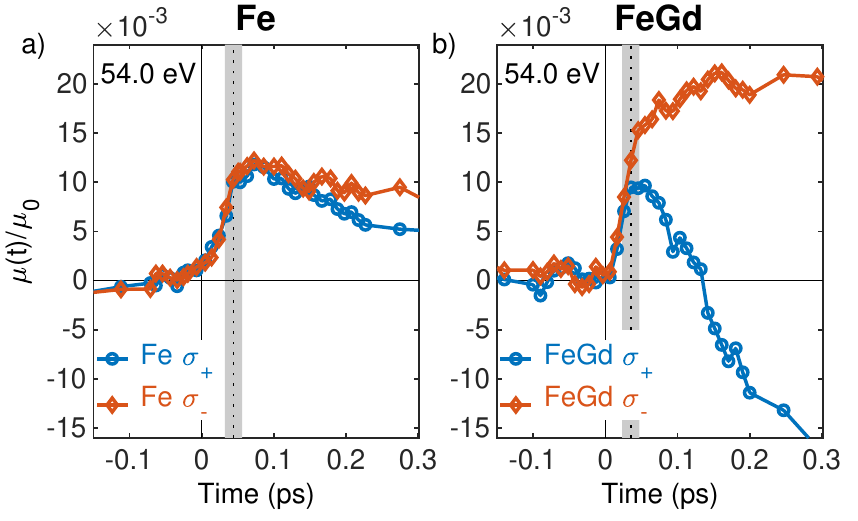}% Here is how to import EPS art
\caption{Normalized helicity dependent absorption $\mu(t)/\mu_0$ for a) a Fe film and b) a GdFe alloy at \SI{54.0}{eV}. The onset of absorption changes and of disparate dynamics in the two channels, $\sigma_\pm$, are delayed by approximately $(45\pm15)\,\si{fs}$, marked as grey bars.}
\label{fig:FeFeGd} 
\end{figure}

We generate light in the extreme ultraviolet spectral range by HHG. Intense laser light pulses with a central wavelength of $\lambda= \SI{800}{nm}$ and a pulse length of $\tau = \SI{25}{fs}$ at a repetition rate of 3 kHz are focused ($f = \SI{1000}{mm}$) into a neon gas cell. The emitted XUV spectrum consists of peaks, spaced by \SI{3.1}{eV} in the energy range between \SI{45}{eV} and \SI{72.5}{eV} with a maximal spectral width of \SI{0.2}{eV}. Fe, Co and Ni films as well as the Ni$_{0.5}$Fe$_{0.5}$ alloy with a thickness of \SI{15}{nm} were magnetron sputtered on thin Si$_3$N$_4$ membranes (\SI{20}{nm}) and exhibit an in-plane magnetization with a coercive field below \SI{10}{mT}. The Gd$_{20}$Fe$_{80}$ alloy was grown on an Al foil (thickness \SI{500}{nm}) with a magnetic easy axis pointing out of the sample plane. Its coercive field is \SI{300}{mT}. 
The experiments were performed in a collinear pump-probe geometry: the samples were excited with femtosecond light pulses with a center wavelength of $\lambda= \SI{800}{nm}$ and a fluence between $\approx 10-15\,\si{mJcm^{-2}}$ and the time delayed, circularly polarized \cite{Vodungbo2011,Willems2015,VonKorffSchmising2017a} XUV radiation was transmitted through the sample, spectrally dispersed by a flat field concave grating and detected with an in-vacuum charged coupled device. For a finite projection of the magnetization and the \textbf{k} vector of the XUV probe radiation, samples with an in-plane/out-of-plane magnetization were mounted at \ang{45}/\ang{90} grazing incident angle. The magnetization direction of the samples was set by an external magnetic field of an electromagnet. An upper boundary of the time resolution is given by an optical cross-correlation of \SI{50}{fs} (full width at half maximum) of the pump-pulse and the residual IR driving pulse of the HHG. For every time step we recorded the transmitted intensity for two opposite magnetization directions, equivalent to an intensity measured with positive and negative helicity, $I(E,t,\sigma_\pm)$, at a fixed magnetization direction. To improve the signal to noise ratio, we acquired the incoming spectrum with a separate spectrometer for normalization \cite{Willems2020} and repeated each time delay scan up to 100 times.
\section{\label{sec:Results}Results:\protect}
We calculate the energy and helicity dependent transient absorption as $\mu(t,\sigma_\pm)= 1-I(t,\sigma_\pm)$, neglecting the small specular reflectance in the XUV spectral range. We attribute changes of the absorption to changes of the electronic system.  This assumption was corroborated experimentally by confirming that the average of $\mu(t,\sigma_\pm)$ corresponds to the response measured with linearly polarized light (cf. Supplementary Material \footnote{See Supplementary Material for additional data acquired with linearly polarized light}). We display the normalized absorption for two opposite magnetization directions measured in the vicinity of the respective $M_{2,3}$ resonances in Fig. \ref{fig:CoNi} for Co and Ni and in Fig. \ref{fig:FeFeGd} for Fe and GdFe. The resonant photon energies at which the absorptive part of the dichroic index of refraction, $\Delta\beta$, is at a maximum are \SI{55.0}{eV} for Fe, \SI{60.1}{eV} for Co and \SI{66.6}{eV} for Ni \cite{Willems2019} (cf. Fig. \ref{fig:disscussion}). They, therefore, do not fully coincide with the photon energies of the equally spaced high harmonic emission peaks. While we probe Co slightly above its $M_{2,3}$ resonance, Fe and Ni is measured \SI{1.0}{eV} and \SI{0.6} eV below the maximum, respectively. In Fig. \ref{fig:CoNi} b) and d), we additionally show the magnetic asymmetry calculated according to
\begin{equation}
\label{eq:asymmetry}
A(t) =\frac{ I(t,\sigma_-)-I(t,\sigma_+)}{I(t,\sigma_-)+I(t,\sigma_+)} \propto M(t) \propto \Delta\beta(t),
\end{equation}
where $\Delta\beta$ is the absorptive part of the magneto-optical function \cite{Willems2019}. Equation \ref{eq:asymmetry} implies that changes in the magnetization are zero when $dI(t,\sigma_-)/dt=dI(t,\sigma_+)/dt$, which is important for the following discussion. At the Co resonance (Fig. \ref{fig:CoNi} a)) we observe distinctly different dynamics in the two absorption channels, $\sigma_\pm$, at $t=0$ and a concomitant drop of the magnetization. Very differently for Ni (Fig. \ref{fig:CoNi} c)), the absorption of both channels shows an identical relative increase up to approximately \SI{55}{fs} before they show a disparate evolution. Consequently, the drop in the \textit{magnetic} asymmetry, cf. Eq~\ref{eq:asymmetry}, shows a delayed onset  as compared to the start of the electronic dynamics at $t=0$. We describe the measured asymmetry by a single-exponential decay function convolved with a Gaussian function (FWHM = \SI{50}{fs}) yielding $t_0 = (0\pm10)\,\si{fs}$ and $t_0 = (55\pm12)\,\si{fs}$ for Co and Ni, respectively. As seen in Fig. \ref{fig:FeFeGd}, we observe a similar behavior for Fe and GdFe. At the resonance the absorption in both channels increases simultaneously and only after about $(45\pm15)\,\si{fs}$ a different temporal evolution in the two absorption channels starts, again leading to a delayed response of the deduced magnetic asymmetry signal. 

\section{\label{sec:Discussion}Discussion:\protect }
\begin{figure}
\includegraphics{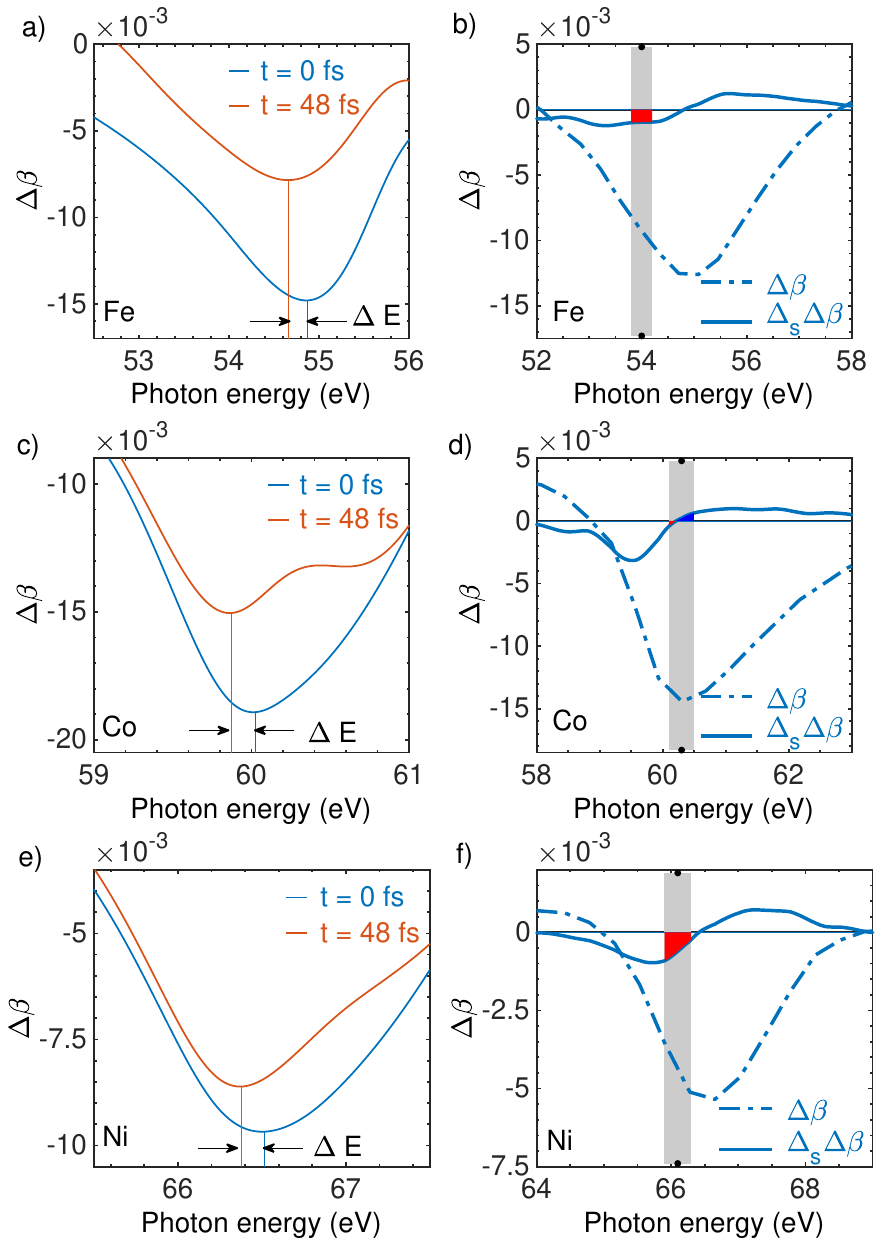}% Here is how to import EPS art
\caption{Calculation of the absorptive part of the magneto-optical functions $\Delta\beta(E)$ for a) Fe, c) Co and e) Ni before and after optical excitation. We observe a reduction of the amplitude of $\Delta\beta$ and a spectral redshift. Panels b), d) and f) show measured values of $\Delta\beta(E)$ (dashed lines) \cite{Willems2019} and the difference spectra, $\Delta_s(\Delta\beta)$, after shifting by $\Delta E = \SI{0.15}{eV}$ to lower photon energies while keeping the amplitude of $\Delta\beta$ constant (solid lines). The red/blue areas indicate an increase/decrease of the MCD signal at the probing photon energies.}
\label{fig:disscussion}
\end{figure}

The observation of a delayed onset of the electronic and magnetic response is unexpected and has not been previously reported in the literature. The origin must either be related to a microscopic physical mechanism leading to a delayed ultrafast change of the magnetic moment or be specific to details of the XUV probing mechanism. The first hypothesis that the magnetization remains unchanged for up to \SI{55}{fs} after an intense optical excitation seems very unlikely due to several reasons. First, experiments providing an absolute temporal resolution of $\lesssim \SI{5}{fs}$ with optical \cite{Shokeen2017a} or XUV \cite{Siegrist2019} spectroscopy have not reported on any delay of the ultrafast demagnetization. Second, early time magnetization dynamics in single element systems is directly related to spin-orbit coupling with an intrinsic time scale given by the interaction strength corresponding to $\lesssim \SI{15}{fs}$ \cite{Elliott2020}. Therefore, corresponding theoretical data, calculated by time dependent density functional theory \cite{Krieger2015}, does not predict any delay between electronic and magnetic dynamics after an applied pump laser pulse.  Furthermore, non-local effects \cite{Battiato2010} which may be active on this time scale as well, can be neglected because we average over the depth of our samples in a transmission experiment. The very different temporal evolution after optical excitation between the individual elements Co, Ni and Fe is another strong indication that a fundamental microscopic origin is doubtful.

Instead, we explain our observations qualitatively by postulating shifts of the absorption and MCD spectra invoked by the generation of a non-equilibrium electronic distribution. Laser excitation with photons of $\approx \SI{1.5}{eV}$ energy will result in an ultrafast change of electron occupations, emptying states below and filling previously available states above the Fermi energy. As our probing photon energies of \SI{60.3}{eV} for Co, \SI{66.1}{eV} for Ni and \SI{54.0}{eV} for Fe are positioned at or below the maximum of the respective resonances (cf. Fig. \ref{fig:disscussion}), optical transitions to suddenly available states below the Fermi level after $t=0$ become allowed. Then, if changes of the absorption can be predominantly attributed to changes of the electronic occupations, additional transitions become available and result in the initial absorption increase around the $M_{2,3}$ resonance photon energies. The consecutive decrease of absorption in the $\sigma_+$ channel is attributed to an increase of minority carriers due to spin-flip scattering during the demagnetization process \cite{Willems2020}. In other words, empty states below the Fermi energy effectively entail a redshift of the absorption spectrum.  Data recorded \SI{3}{eV} above the resonance exhibits an ultrafast decrease of the absorption (cf. Supplementary Material \footnote{See Supplementary Material for additional data acquired at off-resonant photon energies as well as for a more detailed analysis of a transiently shifted absorption spectrum.}) providing further support of the described scenario. 

Since spin-flip processes, i.e., changes between minority and majority carriers, have been postulated to happen around the Fermi energy, where the phase space for scattering events is largest \cite{Carva2011a,Dewhurst2018a}, a non-equilibrium electron distribution inherently leads to a shift of the MCD spectrum as well. 
To corroborate this hypothesis we have performed \textit{ab-initio} calculations of the absorptive magneto-optical function $\Delta\beta(E)$ after laser excitation \cite{Dewhurst2020} and show the result for exemplary demagnetization amplitudes for Fe, Co and Ni in Fig. \ref{fig:disscussion} a), c) and e). We observe that the decrease in amplitude of $\Delta\beta$, shown at $t=\SI{48}{fs}$, is accompanied by a shift as well as by small changes of the spectral shape after the optical excitation. We read off a redshift between the ground and excited state at $t=\SI{48}{fs}$ of $\Delta E \approx \SI{0.15}{eV}$ for all three elements. The calculations reveal that the absolute amplitude of the shift depends sensitively on the number of excited electrons as well as the relative demagnetization amplitude.

To gain a \textit{qualitative} understanding what the assumption of a redshift of the MCD spectrum implies for time resolved MCD spectroscopy employing probe pulses with a finite bandwidth, we have plotted  $\Delta\beta(E)$ \cite{Willems2019} (dashed-dotted lines) as well as $\Delta_s\Delta\beta(E) = \Delta\beta(E-0.15\,\mathrm{eV})-\Delta\beta(E)$ (solid lines) as a function of photon energy for Fe, Co and Ni in Fig. \ref{fig:disscussion} b), d) and f). The grey bars represent the probing high harmonic emission peaks with a spectral width of \SI{0.2}{eV}. We can now appreciate that a spectral shift or spectral reshaping of $\Delta\beta$ will lead to an MCD observable, that sensitively depends on the probing photon energy. For Co, we probe at the zero crossing of $\Delta_s\Delta\beta$ and hence expect a minimal influence to the magnetization dynamics. Very differently, for Fe and Ni, where we probe below the maximum of $\Delta\beta$, a shift leads to an increase of the dichroic response (cf. red areas in Fig. \ref{fig:disscussion} b),d) and f)). Now, we understand the apparent delay between the electronic and magnetic response as the result of two competing processes. Namely the increase of the MCD signal due to a redshift and on the other hand the decrease of the MCD signal due to ultrafast demagnetization.

An ultrafast energy shift of the Ni $L_3$ absorption edge on the order of $\approx \SI{0.15}{eV}$ was previously observed  and attributed to a shrinking of the valence band due to increased localization of the $3d$ bands, however, no shift of the MCD spectrum was detected \cite{Stamm2007a,Kachel2009}. On the other hand, a recent joint experimental and theoretical study revealed pronounced shifts of the calculated off-diagonal dielectric tensor at the $M_{2,3}$ resonance depending both on the specific element as well as on the type of assumed microscopic demagnetization channels \cite{Jana2020}. While the idea that the transient shape of dichroic absorption spectra may reveal details about the microscopic origin of different types of spin excitations has been predicted theoretically some time ago \cite{Erskine1975}, further experimental and theoretical research is required to allow a fully quantitative analysis.

It is important to realize, that the hypothesis of a spectral shift of the MCD spectra suggests that measurements of early time magnetization dynamics require either a broadband radiation source, or a careful adjustment of the photon energy to the maximum dichroic contrast to minimize effects of transiently reshaping absorption spectra. Note that this will become increasingly challenging in experimental geometries like small angle X-ray scattering or coherent imaging which are sensitive to both the real and imaginary parts of the dichroic index of refraction and which may require narrow bandwidth pulses for an optimized spatial resolution.  
\begin{figure}
\includegraphics{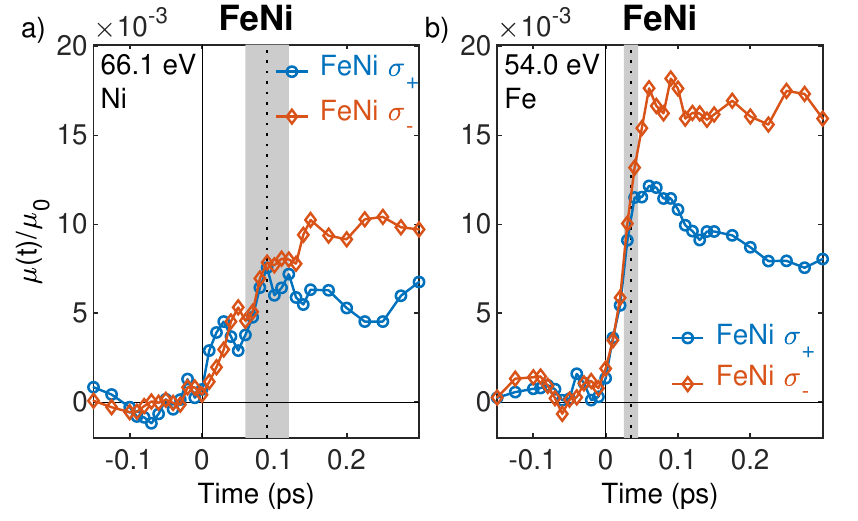}% Here is how to import EPS art
\caption{ Normalized helicity dependent absorption $\mu(t)/\mu_0$ for a NiFe alloy at \SI{66.1}{eV} a) and \SI{54.0}{eV} b) corresponding to the $M_{2,3}$ resonances of Ni and Fe, respectively. We observe a delayed onset of the magnetic asymmetry with respect to changes in the absorption of about $(100\pm30)\si{fs}$ for Ni and $(35\pm12)\si{fs}$ for Fe.}
\label{fig:FeNi}
\end{figure}

Finally, we like to draw the attention to Fig. \ref{fig:FeNi}, which displays the helicity dependent absorption for a Fe$_{0.5}$Ni$_{0.5}$ alloy. Again, we observe an identical increase of both absorption channels at very early times and a disparate evolution after approximately $(35\pm12$)\,\si{fs} and $(100\pm30$)\,\si{fs} for Fe and Ni, respectively. This observation is interesting in view of the intensely discussed temporal delay between the onset of the magnetic response between Fe and Ni in FeNi alloys \cite{Mathias2012,Radu2015,Eschenlohr2017,Jana2020}. Note that such a delay has been very recently explained by two competing processes acting in this two-component material, namely, ultrafast demagnetization and the microscopic process of optical intersite spin transfer, termed OISTR \cite{Hofherr2020}. While our data, suggesting an increased delay for Ni, may be considered as a further support of this explanation, we - nonetheless - like to  point out that when discussing delays in ultrafast magnetic XUV spectroscopy the findings of this communication should be considered.
\section{\label{sec:Conclusion}Conclusion:\protect }
We have presented experimental evidence of a delayed onset of the electronic versus magnetic response in ultrafast resonant magnetic circular dichroism experiments. In our systematic study including the transition metals Fe, Co and Ni as well as the two magnetic alloys FeNi and GdFe, we find delays of up to \SI{55}{fs} for single-element system and up to \SI{100}{fs} in a FeNi alloy, too large to have a true physical origin. Instead, we provide an explanation based on an energy-shift of the absorptive part of the optical as well as magneto-optical functions towards lower photon energies in agreement with state-of-the-art \textit{ab-initio} calculations.  We stress that these findings are of importance for experiments at FEL and high harmonic radiation sources operating with a finite bandwidth in the XUV spectral range which are becoming increasingly common tools for investigations of ultrafast magnetization dynamics.

\section{Supplementary Material}
\begin{figure}[h]
\includegraphics{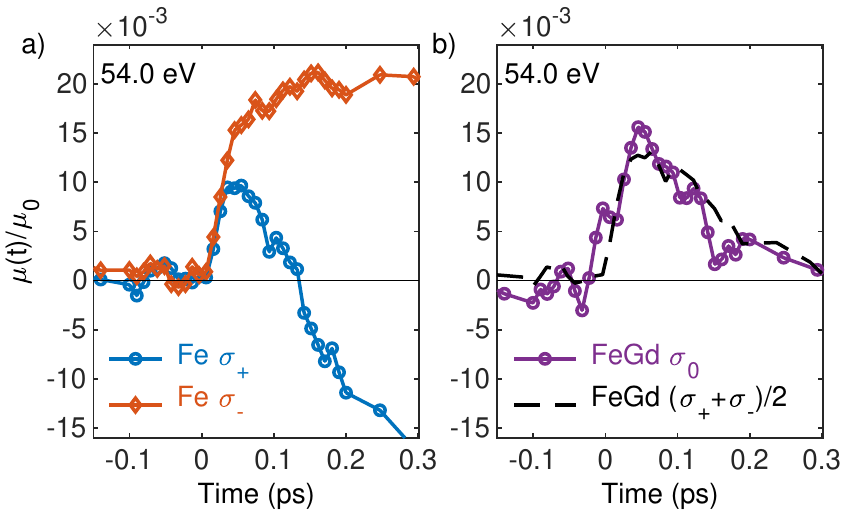}% Here is how to import EPS art
\caption{Normalized absorption $\mu(t)/\mu_0$ for GdFe measured at the $M_{2,3}$ resonance at \SI{54.0}{eV} with a) circularly, $\sigma_\pm$, and b) linearly polarized, $\sigma_0$, XUV photons.  In panel b) we additionally compare results for linearly polarized light with the average of data measured with circularly polarized light (black dashed line) and find an excellent agreement.  }
\label{fig:subFeGdlincirc}
\end{figure}
To further support our assumption that the helicity dependent absorption is a good measure of the electronic response of the magnetic system, we compare our results with measurements performed with linearly polarized XUV radiation, $\sigma_0$. All other parameters of the experiment remained unchanged. Experimental verification corresponds to testing the following identity:
\begin{equation}
\label{eq:asymmetry}
\frac{\mu(t,\sigma_0)}{\mu(t<0,\sigma_0)}=\frac{1}{2} \left(\frac{\mu(t,\sigma_+)}{\mu(t<0,\sigma_+)}+\frac{\mu(t,\sigma_-)}{\mu(t<0,\sigma_-)}\right).
\end{equation}
We plot both sides of the equation in Fig. \ref{fig:subFeGdlincirc} b) and find a very good agreement. 
\section{Supplementary data:  transient absorption}
\begin{figure}[h]
\includegraphics{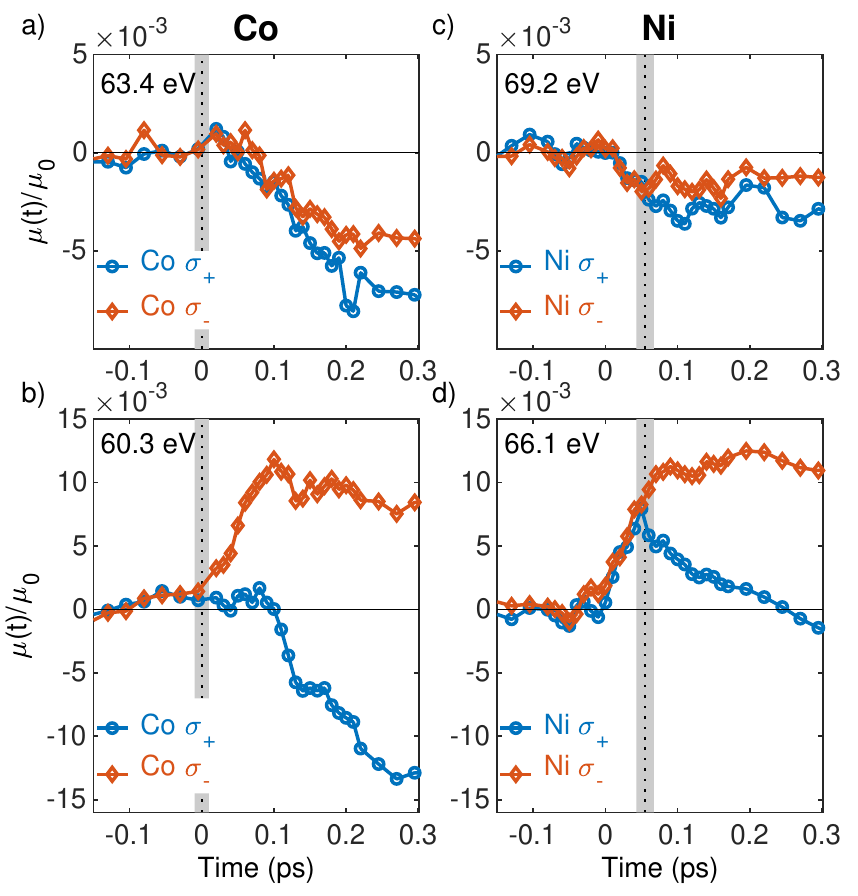}% Here is how to import EPS art
\caption{Normalized helicity dependent absorption $\mu(t)/\mu_0$ close to and \SI{3}{eV} above the respective $M_{2,3}$ resonances for a Co film in panels a) and b) and a Ni film in panels c) and d). For the photon energies, which probe states above the Fermi energy, we observe an ultrafast drop in both channels of the helicity dependent absorption. }
\label{fig:subCoNi}
\end{figure}
\begin{figure}
\includegraphics{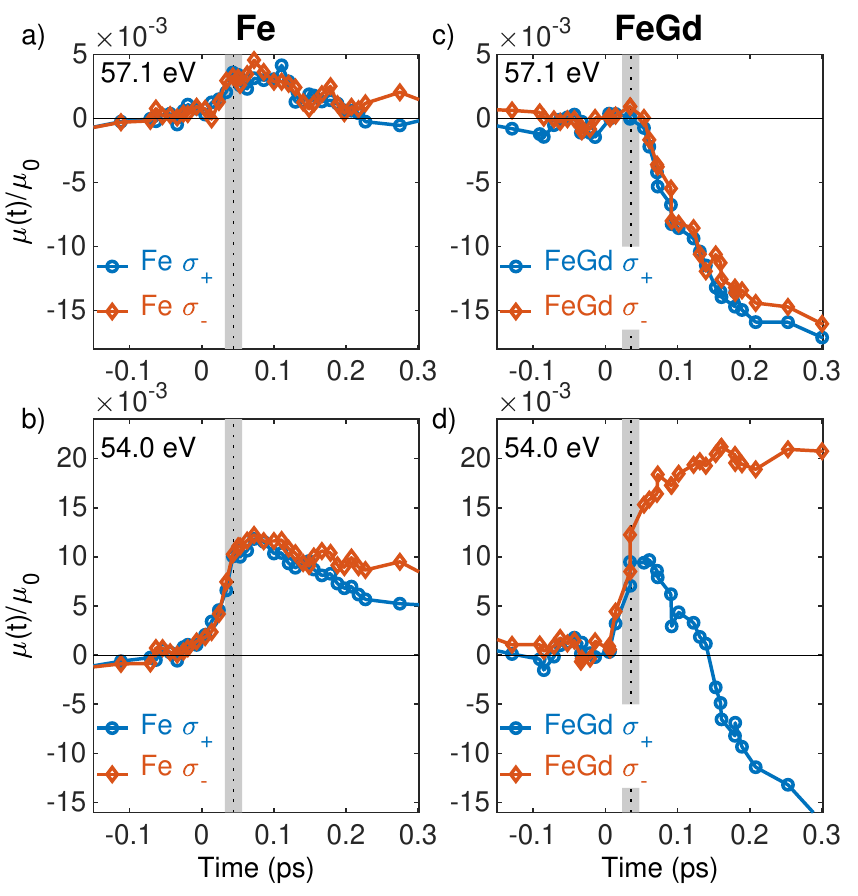}% Here is how to import EPS art
\caption{Normalized helicity dependent absorption $\mu(t)/\mu_0$ measured at \SI{57.1}{eV} and at the resonance at \SI{54.0}{eV} for a Fe film in panels a) and b)  and a GdFe alloy in panels c) and d). For photon energies \SI{3}{eV} above the resonance the absorption exhibits a small increase for Fe and a decrease for GdFe. }
\label{fig:subFeFeGd}
\end{figure}

In the following, we provide additional experimental data to further corroborate our hypothesis that the absorption spectra are subject to a redshift after optical excitation. 
Figure \ref{fig:subCoNi} and Fig. \ref{fig:subFeFeGd} display the normalized helicity dependent absorption $\mu(t)/\mu_0$ for Co and Ni as well as for Fe and GdFe, respectively. Compared to Fig. 1 and 2 of the main text, we additionally plot the measured data \SI{3}{eV} above the respective $M_{2,3}$ resonances. 

At or close to the resonance, we observe that for early times the absorption at the resonances increases in both helicity dependent absorption channels. Only once a disparate evolution starts, $\mu(\sigma_+)$ drops, which we attribute to the filling of minority states due to spin-flip scattering events during the demagnetization process \cite{Willems2020}. Differently, \SI{3}{eV} above the resonance the absorption decreases. Here, the two absorption channels $\mu(\sigma_\pm)$ show a comparable or even identical evolution, because $\Delta\beta$ is or approaches zero. As described in the main text, this general behavior is consistent with a non-equilibrium electron distribution after optical excitation. Emptying of states below the Fermi level facilitates additional transitions leading to an increase of the absorption close to the $M_{2,3}$ resonance. Analogously, previously empty states above the Fermi level become occupied after optical excitation, leading to fewer allowed transitions and a drop of the absorption for photon energies above the resonances. 

\newpage
\section{Supplementary Analysis: Energy shift of absorption spectra}
\begin{figure}[h]
\includegraphics{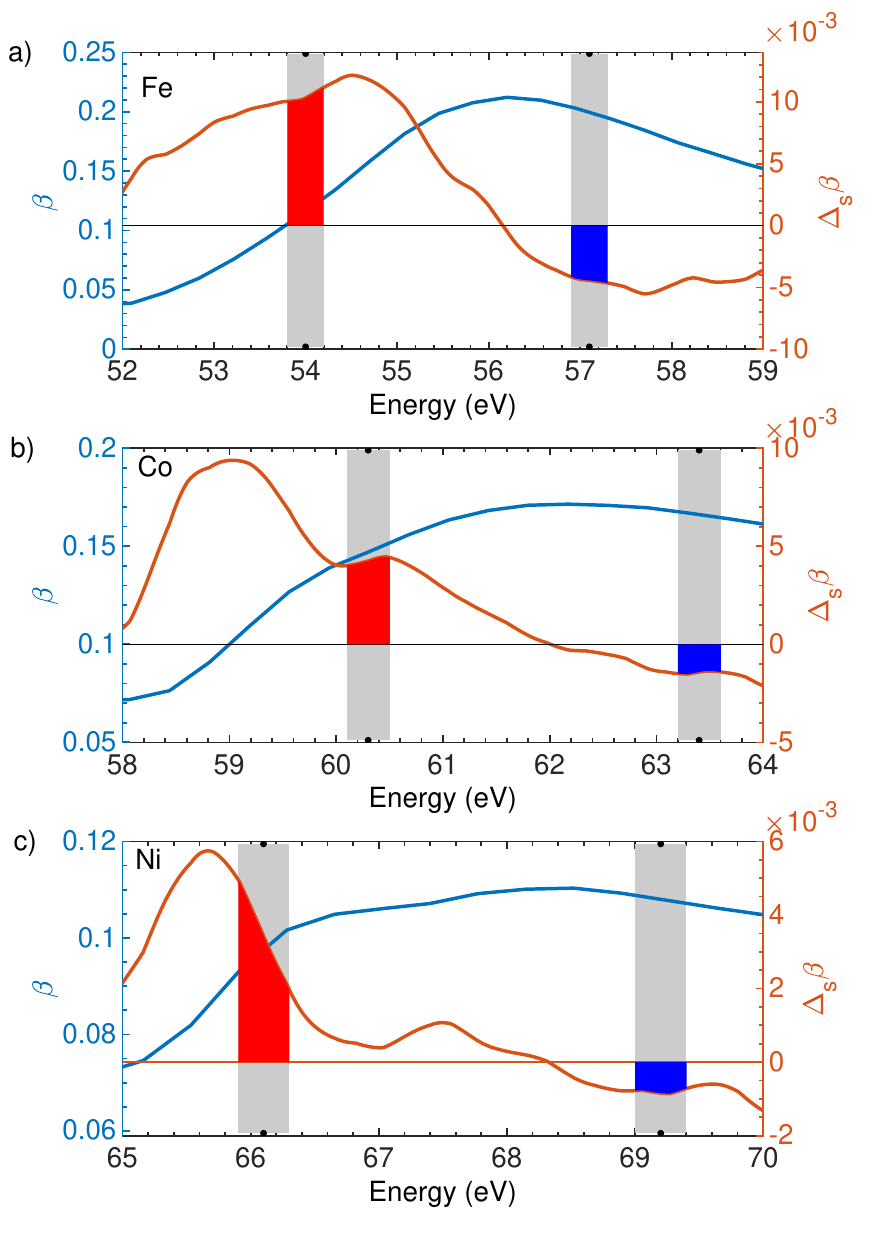}% Here is how to import EPS art
\caption{Absorptive part of the electro-optical functions $\beta$ for Fe, Co and Ni \cite{Willems2019} and changes after shifting by $\Delta E=\SI{0.15}{eV}$ to lower photon energies, $\Delta_s\beta = \beta(E-\Delta E)-\beta(E)$. The grey bars present the probing high harmonic peaks with a width of \SI{0.2}{eV}. The red areas below $\Delta_s\beta$ show the expected absorption changes due to the redshift, positive at the resonance and negative \SI{3}{eV} above the resonance of the respective elements. }
\label{fig:subbetabetashift}
\end{figure}

%Here, we provide an additional illustration on how a spectral line shift effects the absorption measurements performed with narrow bandwidth radiation. To this end, we have plotted measured values of $\beta$ (blue, solid lines) for Fe, Co, and Ni \cite{Willems2019} in Fig. \ref{fig:subbetabetashift} together with the photon energies of the high harmonic peaks close to and \SI{3}{eV} above the respective elemental resonances. The width of peaks in \SI{0.2}{eV} and depicted as grey bars. These energies correspond to \SI{54.0}{eV} and \SI{57.1}{eV} for Fe, \SI{60.3}{eV} and \SI{63.4}{eV} for Co and \SI{66.1}{eV} and \SI{69.2}{eV} for Ni. We then calculated $\Delta_s\beta = \beta(E-\Delta E)-\beta(E)$ for Fe, Co and Ni, using an identical value $\Delta E = \SI{0.15}{eV}$ of a redshift and show the results as red, solid lines in Fig. \ref{fig:subbetabetashift}.
Here, we provide an additional illustration on how a spectral line shift affects the absorption measurements performed with narrow bandwidth radiation. Figure \ref{fig:subbetabetashift} displays both the measured absorption spectra \cite{Willems2019}, $\beta(E)$ (blue solid lines), and the difference spectra, $\Delta_s\beta = \beta(E-\Delta E)-\beta(E)$ (red solid line) with $\Delta E = \SI{0.15}{eV}$ for Fe, Co, and Ni at their respective $M_{2,3}$ resonances. Additionally, we have marked the spectral positions of the \SI{0.2}{eV} broad high harmonic peaks with grey bars. These energies correspond to \SI{54.0}{eV} and \SI{57.1}{eV} for Fe, \SI{60.3}{eV} and \SI{63.4}{eV} for Co and \SI{66.1}{eV} and \SI{69.2}{eV} for Ni.
We can now appreciate that such an assumed redshift entails an increase of absorption for photon energies tuned to the resonance and a decrease of the absorption for photon energies above the resonance. At the probing photon energies we mark the expected increase and decrease due to the shift of the absorption edge as red and blue areas, respectively. Note again that the absolute values of an absorption change will depend on the exact position and width of the probing photon energy, the values of the transient shift as well as on a potential reshaping of the transient absorption spectrum, $\beta(E,t)$. While this makes statements about absolute absorption changes very challenging, we note that the calculated values of $\Delta_s\beta$ not only show a qualitative agreement with our data of $\mu(t)/\mu_0$, but are also on the correct order of magnitude.  

Finally, we like to comment on the off-resonant data for Fe at \SI{57.1}{eV} shown in Fig. \ref{fig:subFeFeGd} a). Here, the absorption increases, which is not consistent with the above explanation and is currently not fully understood. 
Interestingly, a recent study also observed unexpected signal changes for Fe around 57 eV in T-MOKE measurements and discussed this with regard to nonlinearities in the magneto-optical response \cite{Jana2020}. Since such spectral details of the magneto-optical response in the XUV spectral region may also serve as a fingerprint to distinguish different microscopic mechanisms leading to ultrafast demagnetization, e.g., Stoner vs. spin wave excitations \cite{Erskine1975,Turgut2016}, one can expect that further experimental and theoretical efforts in this direction will be very fruitful.

\begin{acknowledgments}
C. v. K. S., S. S. and S. E. would like to thank DFG for funding through TRR227 projects A02 and A04. I. R. acknowledges funding from BMBF through project 05K16BCA and TRR227.
\end{acknowledgments}

K. Y. and F. W. contributed equally to this work. F. W., I. R., C. v. K. S. and S. E. conceptualized the work; K. Y.,  F. W. and C. S. conducted the experiments together with D. S. and C. v. K. S., D. E. and A. T. grew the samples, S. S. and J. K. D. performed the theoretical calculations, F. W., S. S., C. v. K. S., I. R., K. Y. and S. E. discussed and interpreted the results. C. v. K. S. wrote the manuscript with contributions from all authors.
\nocite{*}

%\bibliography{references.bib}% Produces the bibliography via BibTeX.

\end{document}